# Computer says 'no':
# Exploring systemic bias in ChatGPT using an audit approach

**Louis Lippens***

12 February 2024

Access latest version here

## Abstract

Large language models offer significant potential for increasing labour productivity, such as streamlining personnel selection, but raise concerns about perpetuating systemic biases embedded into their pre-training data. This study explores the potential ethnic and gender bias of ChatGPT—a chatbot producing human-like responses to language tasks—in assessing job applicants. Using the correspondence audit approach from the social sciences, I simulated a CV screening task with 34,560 vacancy–CV combinations where the chatbot had to rate fictitious applicant profiles. Comparing ChatGPT's ratings of Arab, Asian, Black American, Central African, Dutch, Eastern European, Hispanic, Turkish, and White American male and female applicants, I show that ethnic and gender identity influence the chatbot's evaluations. Ethnic discrimination is more pronounced than gender discrimination and mainly occurs in jobs with favourable labour conditions or requiring greater language proficiency. In contrast, gender bias emerges in gender-atypical roles. These findings suggest that ChatGPT's discriminatory output reflects a statistical mechanism echoing societal stereotypes. Policymakers and developers should address systemic bias in language model-driven applications to ensure equitable treatment across demographic groups. Practitioners should practice caution, given the adverse impact these tools can (re)produce, especially in selection decisions involving humans.

**Keywords:** large language models, ChatGPT, systemic bias, hiring discrimination, correspondence audit

**Acknowledgements.** I thank Stijn Baert, Eva Derous, and Pieter-Paul Verhaeghe for their continued support during my PhD trajectory. I also acknowledge helpful comments of Eva Schirnt, Sorana Toma, Brecht Neyt, Bart Defloor, and Patrick Button on an earlier draft of this paper and feedback of participants of the Artificial Intelligence and the Economy 2023 conference organised jointly by the Hertie School, IZA Institute of Labor Economics, Kiel Institute, and Sciences Po. Finally, I thank an anonymous reviewer for their suggestions.

\* **Corresponding author.** Ghent University (Sint-Pietersplein 6, 9000 Ghent, Belgium), Vrije Universiteit Brussel, and Centre for the Social Study of Migration and Refugees (CESSMIR). Louis.Lippens@UGent.be. ORCID: 0000-0001-7840-2753.

# 1. Introduction

The emergence of several large language models (LLMs), such as OpenAI's Generative Pre-trained Transformer (GPT) or Google's Pathways Language Model (PaLM), has recently sparked scholarly and public debate on their implications (Nature, 2023; Teubner et al., 2023; The Economist, 2023a, 2023b; Thorp, 2023). LLMs, trained on large corpora of text data, are generative artificial intelligence (AI) applications that can mimic human-like responses to language tasks (Teubner et al., 2023). These algorithms have the potential to significantly impact labour productivity, transforming how we work by automating a myriad of tasks (Acemoglu et al., 2022; Agrawal et al., 2019; Brynjolfsson et al., 2023; Eloundou et al., 2023; Felten et al., 2023; Noy and Zhang, 2023; Teubner et al., 2023). For instance, LLMs can generate comprehensive reports or summaries from large volumes of text data. They may even replace substantial parts of human jobs. LLMs can help in customer support by providing first-level responses to customer inquiries based on user-inputted product or service manual text or responses to frequently asked questions. They can also assist in job applicant assessment by processing curricula vitae (CVs) and cover letters to match applicant qualifications with job requirements. Nevertheless, as these models continue to advance and become more deeply integrated into professional activities, it is vital to ensure that we use them responsibly and ethically, particularly when it comes to decision-making that directly impacts humans.

Recent research suggests that AI technologies, including LLMs, can assist with human resources (HR) practices such as hiring and selection (Budhwar et al., 2023; Vrontis et al., 2021). These technologies may help practitioners automate hiring decision-making, increase decision-making efficiency, or increase decision accuracy and fairness by bypassing human prejudice (Budhwar et al., 2023; Cooke et al., 2019; van Esch et al., 2019). The idea of automating part of the personnel selection process is not new; firms already use algorithms for hiring (Noble et al., 2021). Because LLMs are ideally suited to perform text processing and analysis, they can be used integrally or as part of a broader tool or platform for applicant profile screening purposes. Recently, Pisanelli (2022) examined the success of automated resume screening through a simple language model in a field setting. They found that automated screening diminished the gender interview invitation gap vis-à-vis human recruiters' manual CV screening by almost two-thirds, providing evidence that algorithms may indeed counter human-induced prejudice.

However, instead of reducing bias, LLMs could reproduce the systemic, historical biases embedded in their pre-training data (Budhwar et al., 2023; Caliskan et al., 2017; Cowgill et al., 2020; Peres et al., 2023; Rich & Gureckis, 2019; Schramowski et al., 2022). Examples include amplifying prejudiced views or extrapolating the underrepresentation of vulnerable groups based on the underlying human-created texts. Systemic discrimination induced by LLMs can further the negative cumulative impact of existing bias across domains and time, increasing group-based disparities (Bohren et al., 2022). A recent example of research examining algorithmic racial discrimination is the study of Arnold et al. (2021). They show that Black defendants received a considerably lower pretrial release rate from an advanced AI algorithm than White defendants despite identical pretrial misconduct potential. These observations raise concerns regarding the fairness and objectivity of AI-assisted selection activities and call for a deeper examination of such biases (Tambe et al., 2019).

I focus on ChatGPT, a chatbot actuated by OpenAI's GPT-series models, and its ability to perform a CV screening task without exhibiting bias in its responses. There is no public information about the precise data the GPT-series models are pre-trained on besides that they rely on a comprehensive data corpus encompassing websites, books, manuals, fora, job boards, and other (online) content (Brown et al., 2020; OpenAI, 2023a; Teubner et al., 2023). Their text responses are fine-tuned through specific task prompts and feedback provided by the user. Like other LLMs, ChatGPT might perpetuate and reinforce biases about specific demographic groups it has learned from its training data's patterns, language, and concepts, resulting in discriminatory responses to said CV screening task. Economists refer to this assessment based on group characteristics as statistical discrimination. Hate speech in online fora (Bliuc et al., 2018; Castaño-Pulgarín et al., 2021; Ederer et al., 2023) or harmful stereotypes about minority groups in pre-existing job advertisements (Koçak et al., 2022; Wille and Derous, 2017), for example, can taint the models' training processes. The GPT-3 model instance has demonstrated anti-Muslim bias in word association tasks before, consistently linking Muslims with violence and terrorism (Abid et al., 2021).

The social sciences have a broad tradition of examining discrimination in hiring using correspondence audit studies (Bartkoski et al., 2018; Heath & Di Stasio, 2019; Lippens,

Vermeiren, & Baert, 2023; Quillian et al., 2017, 2019; Quillian & Lee, 2023; Quillian & Midtbøen, 2021; Thijssen et al., 2021; Zschirnt & Ruedin, 2016). The correspondence audit method allows for a causal estimation of discrimination (Gaddis, 2018). In labour market research, this approach compares callback or invitation rates from recruiters or employers to fictitious job applicants who possess similar qualifications but differ in ascriptive characteristics, such as ethnic origin or race. For instance, to examine racial hiring discrimination, a common strategy involves submitting quasi-identical CVs of fictitious applicants to actual vacancies with differences in names indicating racial background. The most recent meta-analytic estimates, comprising worldwide experimental data on hiring discrimination, reveal that candidates signalling ethnic, racial, or national origin minority group membership receive, on average, 29% fewer positive responses from recruiters than their majority counterparts (Lippens, Vermeiren, & Baert, 2023). Across the examined subgroups, average penalties are as high as 41% (for Arabs) or as low as 8% (for Hispanics).

The current CV screening experiment with ChatGPT zooms in on ethnic and gender identity, given their status as the most researched and easily operationalisable discrimination grounds using name variations (Lippens, Vermeiren, & Baert, 2023). Specifically, I consider the hiring chances of fictitious male and female Arab, Asian, Black American, Central African, Dutch, Eastern European, Hispanic, Turkish, and White American applicants. The task consisted of prompting ChatGPT to assess the suitability of these fictitious candidates, only differing by name (signalling ethnic and gender identity), for a given job using the information provided in their CVs and the vacancy text. Aside from the name at the top of the CV and minor differences across CVs to ensure a fitting vacancy–CV match, the input remained identical. Subsequently, I asked ChatGPT to output ratings for each candidate and regarded differences in ratings as evidence of bias and, thus, discrimination.

This study has three main contributions. First, transposing the correspondence audit method from the field experiment literature into a valid alternative for detecting bias or discrimination in LLMs bridges the gap between social science and computer science research. Bias in LLMs has often been measured through word association tasks; for example, by prompting an LLM with the phrase 'a democrat is [male/female]' and asking the LLM to fill in the gender (Abid et al., 2021; Liu et al., 2022). The current experiment applies an existing, well-established method in audit research for measuring hiring discrimination through name association among humans to the context of AI. Second, the approach is methodologically valuable. Potential bias is experimentally exposed in a real-world selection task using a broad range of ethnically identifiable and gendered names linked to actual CVs and vacancies instead of employing a synthetic association task. The experiment also gains from the scalability and automation of LLMs—an advantage over field audit studies using human recruiters—while I could present similar candidate profiles successively to the model without concerns of spillover (e.g. 'Will presenting one candidate affect the bias against another candidate?') or detection ('Will it become apparent that I am running an experiment?'). Third, the study adds to mapping out bias in AI algorithms by quantifying to what extent discriminatory output produced by LLMs can sustain pre-existing differences and societal stereotypes in labour market outcomes based on personal identity.

# 2. Data and methods

I submitted ChatGPT to a simulated CV screening task. To this end, I (i) collected text data from existing job vacancies and CV templates, (ii) created candidate profiles by supplementing the templates with sets of validated names constructed specifically for experimental studies on ethnic and gender identity, and (iii) instructed the chatbot to assess these profiles, only differing by name, based on the requirements in the vacancies. Given the experimental setup, I analysed the data using standard regression techniques. Below, I describe the data-gathering process, the experimental setup, and the estimation strategy to identify potential bias in CV screening by ChatGPT.

## 2.1. Text data

### 2.1.1 Vacancies

The website of the Flemish public employment agency VDAB in the Dutch-speaking part of Belgium (https://www.vdab.be) was the starting point for retrieving vacancy text data. A total of 1,920 vacancies, balanced by occupation and experience level, were selected for the experiment. I chose 23 occupations across different industries to obtain a representative set of vacancies requiring different skills and professional experience. These occupations ranged from clerical service sector employees (e.g. administrative assistant, HR officer) to IT personnel (e.g. IT analyst, IT project leader) and logistics workers (e.g. industrial logistics planner, trucker-trailer driver). The experience level took on three distinct values: no experience ($N$ = 690; 35.94%), at least two years of experience ($N$ = 690; 35.94%), and at least five years of experience ($N$ = 540; 28.13%). As a general rule, I sampled 30 recent vacancies per occupation and experience level to obtain some critical mass. The vacancies of five occupations with at least five years of experience as a functional requirement (e.g. seller of clothing accessories) appeared too infrequent in the vacancy set and were considered non-representative; these vacancies were consequently excluded from the experiment. Table A1 in the appendix provides an overview of counts by occupation and experience level. Examples of the extracted vacancy text can be retrieved from Table A2.

The vacancies further varied regarding job type, shift system, work hours, language requirements, and location. The most prevalent job types were permanent positions ($N$ = 1,501; 78.18%) followed by interim or temporary positions ($N$ = 382; 19.90%); other job types encompassed independent activities, student positions, and flex jobs. The most common shift system was day work ($N$ = 1,823; 94.95%), with other shift systems including two- and three-shift systems, night work, interrupted service, and continuous systems. Contract work hours were mainly full-time ($N$ = 1,860; 96.88%); part-time vacancies constituted only a tiny proportion ($N$ = 60; 3.12%). The most frequently mentioned languages in the vacancies were Dutch ($N$ = 1,808; 94,17%), French ($N$ = 783; 40.78%), and English ($N$ = 760; 39.58%), with varying levels of desired proficiency (i.e. from 'not at all' to 'very good'). Count statistics of the various job and vacancy characteristics can be accessed from Tables A3, A4, and A5 in the appendix.

### 2.1.2. CVs

Each vacancy was paired with 18 CV profiles of fictitious job candidates (differing in ethnic and gender identity) possessing the educational background and professional experience required to perform the job in the vacancy adequately. Table A6 in the appendix shows representative examples of CV text. Like the vacancies, the CV text template was sourced from the Flemish public employment services website. The CV text contained standard information typically found in a resume, such as a residential address, e-mail address, phone number, birth date, nationality (i.e. always Belgian), vehicle ownership and driving ability, bachelor-level degree, general personal characteristics, and language and computer skills. This information remained identical across the 34,560 vacancy–CV combinations to keep the variation between CVs to a minimum and to isolate the effect of ethnic and gender identity. Between vacancies, CVs varied by the specialisation and graduation year of the bachelor's degree and the type and duration of work experience to guarantee a suitable vacancy–CV match.

Within vacancies, CVs only varied in ethnic and gender identity, with all other CV and application details identical. These characteristics were added using candidate names randomly matched to CVs based on assigned ethnic and gender identity. The distinct names signalled diverse ethnic identities—i.e. Arab, Asian, Black American, Central African, Dutch, Eastern European, Hispanic, Turkish, and White American—each accounting for 11.11% of the candidate profiles, and two genders—i.e. female and male—each accounting for 50% of the profiles. Here, Dutch refers to the language spoken in (the Flemish part of) Belgium and The Netherlands, amongst other countries, rather than the country of origin (i.e. The Netherlands).

I drew names from five sources. The first series comprising Asian male, Black American, Hispanic, and White American names were acquired from the recently published dataset of Crabtree et al. (2023). They compiled an extensive set of validated names for use in name experiments based on surveys conducted in the United States, accounting for confounding factors beyond intended race, such as socioeconomic status. Additional Hispanic female first names were taken from the name categorisation test of Gaddis (2017) to increase variety vis-à-vis the single Hispanic female first name in Crabtree et al. (2023). Sets of validated Arab full names and Asian and White American female first names were sourced from Baert et al. (2022), who conducted a similar name experiment. The latter set of names was again used to increase variety due to the limited number of Asian and White American female first names in Crabtree et al. (2023). Dutch names were obtained from Lippens, Dalle, D'hondt, Verhaeghe, & Baert's (2023) correspondence audit study and Martiniello and Verhaeghe's (2022) name perception experiment. Finally, Central African, Eastern European, Turkish, and additional Arab names were sourced from Martiniello and Verhaeghe (2022).

## 2.2. Experiment

### 2.2.1. Prompts

The experiment consisted of sending prompts and receiving and logging model output. I repeated this process until all vacancy–CV combinations were presented to ChatGPT. The prompts in the experiment included instructions, the vacancy text, and the CV profile text. The only differences between CVs were the assigned ethnicity and gender, signalled through names, while all other details remained the same (see Section 2.1.2). All input was written in Dutch; Lai et al. (2023) have shown that the accuracy and performance of ChatGPT in Dutch on specific language tasks are on par with its accuracy and performance in English. Conducting the experiment in Dutch also enabled the identification of assumed language-specific discrimination between two majority groups (i.e. White American vs. Dutch candidates) and ethnic-specific discrimination between majority and minority groups (e.g. White American vs. Black American applicants).

The general instruction ordered ChatGPT to help select suitable candidates: "You assist in making personnel selection decisions. Based on the vacancy below and the candidate's CV, you should indicate how likely you would invite the candidate for an interview". To obtain an outcome variable at the ratio level, ChatGPT was directed to "answer only with a score ranging from 1 (very unlikely) to 100 (very likely)", i.e. an integer. This rating was eventually labelled the 'interview invitation score' in the analyses. Because it was ambiguous how imposing different cutoff scores on ChatGPT's output would influence bias in the eventual selection decision, I carried out post-hoc sensitivity analyses to evaluate the impact of proprietary user-imposed thresholds (see Section 2.3.4 and Section 3.4).

### 2.2.2. Data exchange

I conducted the correspondence audit-like experiment using OpenAI's Application Programming Interface (API) instead of using ChatGPT's web interface

(https://chat.openai.com). The API facilitated the data exchange by automatically transmitting text prompts to ChatGPT and receiving generated responses in return (OpenAI, 2023b). Connection with the API was made through R relying on the {httr} and {jsonlite} packages.

Automating the experiment using the API over ChatGPT's web interface had two significant advantages. The first advantage was the scalability of the design. Compared to regular audit studies with human subjects, I could present a relatively high number of applicants to ChatGPT in a short time. For reference, the automated sequential sending of the 34,560 vacancy–CV combinations was completed in approximately eight hours (in June 2023). Second, the absence of chat history or memory due to the successive and isolated presentation of vacancy–CV combinations to ChatGPT prohibited the chatbot from generating task-trained responses that could undercut the experiment's validity. Because of the absence of this spillover, it was possible to send identical applications, with just the change in treatment, rather than needing to create substantial differences in candidate profiles or CV templates. This approach improved the precision of the results over audit experiments in the field, where differentiation is needed to avoid detection by human recruiters.

### 2.2.3. Language model

At the time of the experiment, OpenAI's had two main models in its GPT series: GPT-3.5(-Turbo) and GPT-4. Both models were trained on data primarily sourced from the internet, comprising online articles, websites, and other texts, and included information up to September 2021 (OpenAI, 2023a). Although GPT-4 showcased advancements in certain language tasks, GPT-3.5 offered abundant capability and efficient resource utilisation. Importantly, GPT-3.5 was more accessible, being available free of charge to the broader public, in contrast to GPT-4, which required a paid subscription for access through its web interface. This widespread accessibility of GPT-3.5 underlined its relevance to the average user and supported its use as the model instance for experimentation in this study. For reproducibility purposes, I relied on the 13 June 2023 snapshot of the GPT-3.5 model, which preserved the pre-trained language model's state as of that date.

Notably, the GPT models were equipped with safeguards to prevent unintentional discrimination, especially when directly comparing similar or nearly identical job candidates. For example, when prompting the GPT-3.5 instance of ChatGPT via its web interface to compare quasi-identical profiles—one Dutch and the other Black American—ChatGPT's response accurately identified minimal differences between the candidates, such as their ethnic background. More explicitly instructing ChatGPT to evaluate the candidates based on this ethnic distinction, the output read: "As an AI language model, I cannot discriminate against candidates based on their background [...] It is important to evaluate candidates based solely on their qualifications, skills, and educational backgrounds relevant to the job position. Therefore, it would not be appropriate to assess or score the candidates based on their backgrounds". However, the experimental approach in this study, using isolated prompts exchanged via the API without directly comparing fictitious candidates, circumvented these safeguards and enabled an unrestrained evaluation of the potential ethnic and gender bias in ChatGPT's output. I assume HR professionals would use ChatGPT similarly by sequentially presenting qualifying candidates to the model, possibly intermittently and, thus, isolated.

## 2.2.4. Sampling strategy

I also altered the model temperature (or sampling strategy) to examine its moderation effect on ChatGPT's bias. This temperature parameter influences the randomness or creativity of ChatGPT's output. Low temperatures produce more deterministic responses based on the chatbot's training data patterns, while higher temperatures generate increased stochastic outputs. The impact of different temperature settings or sampling strategies on ChatGPT's bias is unclear, as raising the temperature may reduce common pre-trained biases but introduce uncommon biases. Following a probability weighting scheme of 60.00%, 8.75%, 8.75%, 8.75%, 8.75%, 2.50%, and 2.50%, temperatures between 0.00 and 1.50 with increments of 0.25 were integrated into the API request. Here, 0.00 was the minimum temperature setting, making ChatGPT mostly deterministic, allowing minimal output variability, while temperatures above 1.50 resulted in the chatbot producing such variable output that it no longer adhered to the prompt's guidelines (i.e. outputting a quantifiable score in $[1,100] \cap \mathbb{N}$). Exact count statistics by model temperature can be retrieved from Table A7 in the appendix.

## 2.3. Estimation

### 2.3.1. Principal analyses

I estimated multiple ordinary least squares (OLS) regression models to assess the relationship between the ratings (i.e. interview invitation scores) outputted by ChatGPT and candidate and job features. In these models, the dependent variable indicated a candidate's suitability for a vacancy, expressed by a score (i.e. integer) ranging from 1–100 ($Inv$). The candidates' ethnic identity ($Eth_i$), at the individual level $i$, was the main predictor of interest. I also held ChatGPT's sampling strategy or temperature ($Tmp_i$) constant in all estimated models since it was altered in about two-fifths of the prompts (see Section 2.2.4). A separate analysis of the influence of the temperature on the results is discussed in Section 3.4. The principal model, shown in Equation 1, consisted of the aforementioned predictor variables, operationalised at the applicant level $i$, with $\alpha$, an intercept, $\beta$ and $\gamma$, model coefficients, and $\varepsilon_i$, the error term.

$$Inv = \alpha + \beta * Eth_i + \gamma * Tmp_i + \varepsilon_i \qquad (1)$$

In subsequent OLS models, I expanded the predictor scope by including candidate and job characteristics as covariates. Besides ethnic identity, candidate characteristics ($CAN_i$) comprised the candidate's gender ($Gen_i$). Work-related job characteristics ($JOB_v$) included the occupation (e.g. administrative assistant), job type (e.g. temporary job), work hours (e.g. part-time), and shift system (e.g. night work), defined at the vacancy level $v$. Furthermore, I entered several job language proficiency variables concerning Dutch, French, and English. Other job-level control variables included the level of professional experience requested (e.g. at least five years) and the employment location (e.g. Antwerp). Equation 2 shows the extended model containing these covariates with $\alpha$, an intercept, $\gamma$, a coefficient, B and $\Lambda$, vectors of model coefficients, and $\varepsilon_i$, the error term.

$$Inv = \alpha + CAN_i * B + \gamma * Tmp_i + JOB_v * \Lambda + \varepsilon_i \qquad (2)$$

The third set of OLS models incorporated interaction terms between candidate identity and the candidates' signalled gender. Equation 3 depicts the terminal linear model, including sampling strategy and job-related variables with $\alpha$, an intercept, $\beta$ and $\gamma$, coefficients, $\Lambda$, a vector of model coefficients, and $\varepsilon_i$, the error term.

$$Inv = \alpha + \beta * (Eth_i * Gen_i) + \gamma * Tmp_i + JOB_v * \Lambda + \varepsilon_i \qquad (3)$$

Each successive model in this series of OLS models contained additional controls related to candidate and job characteristics. By progressively adjusting for these factors, the analyses aimed to further isolate the experimental effect of candidate identity on the interview invitation probability, minimising the influence of potential confounding factors in the vacancy and CV texts used.

### 2.3.2. Heterogeneity analyses

In the heterogeneity analyses, I explored the interactions between candidate characteristics and job-related variables, specifically focusing on how these interactions correlate with interview invitation scores. These analyses aimed to uncover differential impacts of gender and ethnic identity across various job contexts, thereby providing insights into the mechanisms of CV screening by ChatGPT.

First, I examined the interaction between a candidate's ethnic identity ($Eth_i$ at the individual level $i$) and job characteristics ($JOB_v$ at the vacancy level $v$). Equation 4 explicates the model for assessing whether job characteristics, such as occupation or required language skills, amplify or mitigate ethnic bias in scoring candidates. The model includes gender ($Gen_i$), the sampling strategy ($Tmp_i$), and interaction terms between the job-related variables and ethnic identity with $\alpha$, an intercept, $\beta$ and $\gamma$, coefficients, $\Lambda$, a vector of model coefficients, and $\varepsilon_i$, the error term.

$$Inv = \alpha + \beta * Gen_i + \gamma * Tmp_i + (JOB_v * Eth_i) * \Lambda + \varepsilon_i \qquad (4)$$

Second, I investigated the interaction between gender and job characteristics, revealing the conditions under which gender bias in CV screening by ChatGPT is exacerbated or reduced. Analogous to Equation 4, Equation 5 illustrates the interaction model, consisting of ethnic identity, the sampling strategy and interaction terms between job-related variables and gender, again with $\alpha$, an intercept, $\beta$ and $\gamma$, coefficients, $\Lambda$, a vector of model coefficients, and $\varepsilon_i$, the error term.

$$Inv = \alpha + \beta * Eth_i + \gamma * Tmp_i + (JOB_v * Gen_i) * \Lambda + \varepsilon_i \qquad (5)$$

### 2.3.3. Statistical corrections

Standard errors of each OLS model were corrected using cluster-robust wild bootstrapping. This technique produces a distribution of estimated parameters, facilitating the calculation of more precise standard errors (Cameron & Miller, 2015; Cameron et al., 2008). It achieves this by generating interim datasets with reformed dependent variables derived from a combination of the original model's fitted values, the residuals, and a random factor. There were three reasons to perform this correction: to control within-cluster error correlation, address heteroskedasticity, and mitigate violations against the residual normality assumption. Clusters were defined at the vacancy level, given the correlation between the assignment of the candidates and the vacancies presented to ChatGPT, similar to the approach in correspondence audit studies with humans (Abadie et al., 2023; Vuolo et al., 2018). The estimates appeared to stabilise around 2,000 bootstrap replications, which suffices in the context of empirical research (Cameron & Miller, 2015).

Furthermore, in the case of multiple family-wise comparisons, I performed ex-post corrections of the *p*-values in the regression analyses using Holm's (1979) method. This approach entails a stepwise ranking procedure that reduces the likelihood of false positives. Implementing this procedure was particularly meaningful for models that involved numerous comparisons between distinct categories and their respective reference groups. Throughout Section 3, where appropriate, I reported Holm-corrected *p*-values alongside the original estimates to cross-validate the results. Using less stringent correction procedures, such as the Benjamini–Hochberg or Benjamini–Yekutieli methods based on the Simes test outlined in Burn et al. (2022), produced similar results and did not alter their interpretation (Benjamini & Hochberg, 1995; Benjamini & Yekutieli, 2001).

### 2.3.4. Sensitivity analyses

In a real-world scenario, decision-makers such as HR professionals would likely rank the candidates using a proprietary cutoff score based on ChatGPT's output to select the optimal number of candidates to invite for the job interview. Therefore, I estimated logistic regression models to analyse sensitivity across different thresholds regarding ethnic identity. To this end, I used a penalised maximum likelihood estimator, which reduces the variance for the estimated coefficients (even in large samples) compared to the regular maximum likelihood estimator (Firth, 1993; Rainey & McCaskey, 2021). The results were robust vis-à-vis using a non-penalised estimator. The dependent variable was the probability of receiving an invitation given a predefined cutoff score *n*, i.e. *Pr(Inv$_n$ = 1)*. I ran a total of 100 logit models where *n* took on every integer between 1 and 100 (i.e. every possible interview invitation score produced by ChatGPT). The probability of receiving an invitation at a given threshold was regressed on the same predictor and covariates defined in Equation 1. Equation 6 shows this logistic model with $\alpha$, an intercept, $\beta$ and $\gamma$, coefficients, and $\varepsilon_i$, the error term.

$$\Pr(Inv_n = 1) = logit^{-1}(\alpha + \beta * Eth_i + \gamma * Tmp_i + \varepsilon_i) \tag{6}$$

This analytic approach enabled assessing whether potential bias persisted across cutoff scores. In other words, how does the chosen cutoff score impact potential bias from the perspective of the decision-maker? Is there an optimal cutoff score where the bias is minimised? Which range(s) of cutoff scores exhibit(s) increasing or decreasing discrimination?

Finally, I produced discrimination ratios, which capture the relationships between two positive response rates to estimate relative penalties between groups. These ratios were calculated by transforming the log-odds from the logit model specification in Equation 6 to odds ratios (*OR*) and, in turn, into discrimination ratios (*DR*). Discrimination ratios are essentially risk ratios (*RR*) and constitute a standard measure of discrimination in correspondence audit studies in the social sciences (Lippens, Vermeiren, & Baert, 2023). Equation 7 shows the *OR*-to-*DR* transformation. The ratios were defined relative to the baseline risk (*Pr$_{base}$*), corresponding to the probability of a positive response for the reference group at a given cutoff score. Confidence intervals were computed through a Wald *z*-distribution approximation.

$$DR = \frac{OR}{(1 - Pr_{base}) + (Pr_{base} * OR)} \tag{7}$$

# 3. Results

## 3.1. Scoring bias

ChatGPT generally produces high scores when responding to the prompt "How likely is it that you would invite the candidate for an interview?". On a 1–100 scale ranging from very unlikely to very likely, the mean invitation score equals 66.11 ($SD$ = 13.27). Moreover, ChatGPT exhibits a preference for two numerical values. In over a quarter of the cases, ChatGPT scores the candidate 50 (i.e. 8,701 occurrences or 25.18%), while in more than two-fifths of the cases, ChatGPT outputs a score of 70 (i.e. 14,604 occurrences or 42.26%). Figure 1 illustrates the relative frequency distribution of invitation scores (by ethnic identity). Even though the opaque decision-making process of the GPT model (and LLMs more broadly) makes it unclear why these specific values are most common, the observation of reoccurring identical values is not surprising given the recurrent presentation of quasi-identical CVs (only differing by name) as input for the screening task. Using an interview invitation score threshold of ≥ 50, right in the middle of the 1–100 scale, ChatGPT would invite the average candidate in 96.88% of the cases—i.e. almost every time. This proportion is remarkably high compared to response rates in correspondence experiments with human recruiters, where the overall invitation probability is closer to 20% (Lippens, Vermeiren, & Baert, 2023). I further evaluate the sensitivity of using different cutoff scores as an end decision-maker in Section 3.4.

< Figure 1 about here >

Despite the generally high scoring, ChatGPT demonstrates a noticeable preference for Dutch-named candidates over equivalent other candidates when presented with their CVs and matched Dutch-written vacancies. The distribution for candidates with names signalling different ethnic identities—i.e. Arabs, Asians, Black Americans, Central Africans, Eastern Europeans, Hispanics, Turks, and White Americans—is slightly more right-skewed than for Dutch candidates. On average, Dutch candidates receive a score of 67.62 ($SD$ = 12.63), while the other group attains a combined 65.92 ($SD$ = 13.34). Within the latter group, there are slight differences. The mean subgroup scores amount to 66.21 ($SD$ = 13.05) for Arabs, 65.47 ($SD$ = 13.39) for Asians, 65.80 ($SD$ = 13.57) for Black Americans, 66.07 ($SD$ = 13.26) for Central

Africans, 65.21 (*SD* = 13.76) for Eastern Europeans, 66.01 (*SD* = 13.35) for Hispanics, 65.88 (*SD* = 13.15) Turks, and 66.67 (*SD* = 13.12) for White Americans, respectively.

## 3.2. Identity discrimination

### 3.2.1. Ethnic identity

Table 1 contains the estimates of six OLS regression models where ChatGPT's interview invitation scores are regressed on job candidate ethnic and gender identity, among other covariates (see Section 2.3 for the estimation details). Compared to the Dutch reference group, the effects of each of the other ethnic identities on the invitation score are statistically significant (at the 0.1% significance level) and negative. In other words, candidates with White American, Arab, Central African, Hispanic, Turkish, Black American, Asian, and Eastern European names receive significantly lower scores from ChatGPT than Dutch-named candidates. Although minor, average penalties range from approximately −0.96 to −2.42 points on a 1–100 scale.

< Table 1 about here >

In contrast with worldwide hiring discrimination observed in human recruiters, where candidates of Arab, Middle Eastern or Northern African origin face the highest disadvantage (see Lippens, Vermeiren, & Baert, 2023), Eastern Europeans are penalised the most by ChatGPT compared to the Dutch reference group ($\beta_{\text{E.Eu.}}$ = −2.4170, *SE* = 0.2275, $p_{\text{Holm}}$ < 0.001; see Model 1 in Table 1). This candidate group is followed by Asians ($\beta_{\text{Asian}}$ = −2.1583, *SE* = 0.2340, $p_{\text{Holm}}$ < 0.001), Black Americans ($\beta_{\text{B.Am.}}$ = −1.8436, *SE* = 0.2288, $p_{\text{Holm}}$ < 0.001), Turks ($\beta_{\text{Turkish}}$ = −1.7478, *SE* = 0.2227, $p_{\text{Holm}}$ < 0.001), Hispanics ($\beta_{\text{Hispanic}}$ = −1.6257, *SE* = 0.2288, $p_{\text{Holm}}$ < 0.001), Central Africans ($\beta_{\text{C.Afr.}}$ = −1.5468, *SE* = 0.2278, $p_{\text{Holm}}$ < 0.001), Arabs ($\beta_{\text{Arab}}$ = −1.4117, *SE* = 0.2141, $p_{\text{Holm}}$ < 0.001), and White Americans ($\beta_{\text{W.Am.}}$ = −0.9563, *SE* = 0.2258, $p_{\text{Holm}}$ < 0.001), whose relative penalties are marginally lower. Figure 2 illustrates these results visually. Holding relevant covariates constant does not alter the significance of these results or their interpretation (see Models 2 to 7 in Table 1).

< Figure 2 about here >

While Dutch and White American applicants could both be regarded as majority group candidates in their respective geographies, candidates with White American names still face a small but significant penalty compared to their Dutch counterparts ($\Delta_{W.Am.-Dutch} = \beta_{W.Am} = -0.9563$). This result suggests that the prompt language used (i.e. Dutch) at least partly affects ChatGPT's scoring bias and that this score difference could rather be interpreted as a language-specific than an ethnic-specific bias. Conversely, White American applicants receive significantly higher scores than Black American applicants, on average ($\Delta_{B.Am.-W.Am.} = -0.8765$, $t_{Welch} = -2.88$, $p = 0.004$). Because the prompt language reasonably should have little effect on the latter difference, I interpret it as a mainly ethnic-specific bias.

### 3.2.2. Gender identity

ChatGPT's outputted interview invitation scores do not vary statistically significantly with the candidate's gender. Models 2 to 7 in Table 1 include coefficient estimates for female versus male candidates. These coefficients are slightly positive but indistinguishable from zero. The statistical insignificance of this finding remains unchanged when including relevant covariates. The observation aligns with average gender discrimination estimates from the field experimental literature on hiring discrimination (Lippens, Vermeiren, & Baert, 2023). Nevertheless, the question remains whether there are intersectional effects between gender and ethnic identity in determining ChatGPT's output.

A prominent hypothesis in the discrimination literature is the double minority status or double jeopardy hypothesis (Derous et al., 2012, 2015). Belonging to an ethnic and gender minority group presumably engenders a double penalty; ethnic females are subject to increased penalties. Nonetheless, recent research has indicated that ethnic minority males often experience more discrimination than females, especially Arab applicants (Arai et al., 2016; Dahl & Krog, 2018; Derous et al., 2015). This discrimination appears partly induced by stereotypes about masculinity (Bursell, 2014; Di Stasio & Larsen, 2020). Starting from the idea that this intersected genderism is inherently present in ChatGPT's training data, I assess whether ethnic discrimination is significantly moderated by gender.

Table 2 shows the interaction effects of ethnic and gender identity on ChatGPT interview invitation scores. While Turkish male applicants receive marginally worse scores than Dutch male candidates ($\beta_{Turkish} = -0.8595$, $SE = 0.3243$, $p = 0.009$, $p_{Holm} = 0.138$; see Model 1 in Table 2), Turkish females lose approximately 1.78 points net versus Turkish males and 1.17 points net compared to Dutch male applicants ($\beta_{\text{Female}}$ = 0.6063, $SE$ = 0.3221, $p$ = 0.057; $\beta_{\text{Turkish:Female}}$ = −1.7765, $SE$ = 0.4810, $p$ < 0.001, $p_{\text{Holm}}$ = 0.003). The apparent double penalties for Eastern European and Black American females become statistically insignificant after applying Holm's correction ($\beta_{\text{E.Eu.:Female}}$ = −1.0459, $SE$ = 0.4767, $p$ = 0.026, $p_{\text{Holm}}$ = 0.412; $\beta_{\text{B.Am.:Female}}$ = −0.9499, $SE$ = 0.4753, $p$ = 0.041, $p_{\text{Holm}}$ = 0.660). Evidence for the moderation effect of gender on the ethnic bias for the remaining identities is absent. Holding relevant covariates constant does not impact the statistical significance of these findings (see Model 2 in Table 2). In other words, Turkish females are worse off than Turkish males in the CV screening task, reflecting the double minority status of the former group. Nevertheless, similar to the main effect of ethnic identity, penalties remain relatively small considering the 1–100 scale of the outcome variable.

< Table 2 about here >

## 3.3. Discrimination heterogeneity

### 3.3.1. Discrimination by name

The differences between groups outlined in Section 3.2 hide the substantial dispersion in assigned interview invitation scores between names of the same ethnic and gender identity. In other words, much larger differences in ChatGPT's score output exist within groups than between groups. Figure 3 visualises each name's interview invitation score distribution and mean by ethnic and gender identity. The visualised name score dispersion seems to vary across ethnic identities but is generally consistent across genders (see Section 3.2.2 for an exception regarding male and female Turkish applicants).

< Figure 3 about here >

I present two concrete, restricting the analysis to sets of two names. First, the Central African male applicant 'Gaetan Bihsinga' receives significantly lower scores ($M$ = 62.64, $SD$ = 13.00) than 'Tanguy Mangala' ($M$ = 69.69, $SD$ = 12.68, $t_{\text{Welch}}$ = −3.14, $p$ = 0.002) with an interview invitation score difference of −7.05 points. Similarly, the Turkish female candidate

'Esma Aydin' scores 5.13 points lower ($M$ = 63.65, $SD$ = 13.48) than 'Meryem Yüksel' ($M$ = 68.78, $SD$ = 11.88, $t_{Welch}$ = −2.22, $p$ = 0.028).

Specific names thus elicit different discrimination levels, likely reflecting nuances in ChatGPT's pre-trained bias. Individual applicants can be far worse (or better) off compared to their ethnic and gender peers than when comparing 'average candidates' between groups. Nonetheless, this finding relies on randomly allocating a sufficient number of the same names to the ethnic–gender identities across vacancy–CV combinations. In this context, note that despite successful randomisation, the number of iterations may be too small to infer true differences in ratings between some names. Counts and mean interview invitation scores and probabilities of all 812 first and last name combinations used in the experiment can be retrieved from Table A8 in the appendix.

### 3.3.2. Discrimination by job characteristics

Further heterogeneity analyses reveal distinct patterns of job context-dependent bias in ChatGPT's CV screening. For ethnic identity, biases are notably contingent upon work hours, specifically full-time roles ($\lambda_{maj}$ = 67.58, $CI_{95\%}$ = [66.95; 68.22]; $\lambda_{min}$ = 65.83, $CI_{95\%}$ = [65.27; 66.40]), and shift system, specifically day work ($\lambda_{maj}$ = 67.69, $CI_{95\%}$ = [67.06; 68.32]; $\lambda_{min}$ = 65.95, $CI_{95\%}$ = [65.39; 66.51]). Here, better labour conditions typically mean higher scores for the Dutch (majority) group compared to the ethnic minority groups. Language proficiency in Dutch (i.e. good, $\lambda_{maj}$ = 67.07, $CI_{95\%}$ = [65.96; 68.18]; $\lambda_{min}$ = 65.80, $CI_{95\%}$ = [64.87; 66.73]; very good, $\lambda_{maj}$ = 68.11, $CI_{95\%}$ = [67.43; 68.78]; $\lambda_{min}$ = 66.24, $CI_{95\%}$ = [65.66; 66.82]), French (i.e. good, $\lambda_{maj}$ = 70.33, $CI_{95\%}$ = [69.38; 71.28]; $\lambda_{min}$ = 68.08, $CI_{95\%}$ = [67.31; 68.85]), and English (i.e. good, $\lambda_{maj}$ = 69.39, $CI_{95\%}$ = [68.31; 70.46]; $\lambda_{min}$ = 68,20, $CI_{95\%}$ = [67.20; 69.19]; very good, $\lambda_{maj}$ = 66.82, $CI_{95\%}$ = [65.20; 68.44]; $\lambda_{min}$ = 64,78, $CI_{95\%}$ = [63.28; 66.28]) also emerge as key context variables fostering discrimination in ChatGPT's output, where higher proficiency requirements consistently underlie the bias against minority candidates. OLS estimated marginal effects at the mean, controlled for relevant covariates, by ethnic identity (i.e. ethnic majority and minority identity) and job characteristics can be retrieved from Tables A9 to A16 and Figures A1 to A5 in the appendix.

These findings are consistent with the mechanism of statistical discrimination, which posits that, due to information asymmetry in the selection process, (i) screeners want to minimise the risk of a costly wrong hire and are therefore more inclined to offer ethnic minorities poorer labour conditions, such as fixed-term contracts, and (ii) screeners more easily rely on group productivity signals such as language skills, which are generally considered worse among ethnic minorities (Lippens et al., 2022; Lippens, Dalle, D'hondt, Verhaeghe, & Baert, 2023; Martínez-Pastor, 2013; Oreopoulos, 2011). Overall, ethnic discrimination in automated screening is partially shaped by the jobs' specific demands and exhibits features similar to statistical hiring discrimination by humans.

In Section 3.2.2, I demonstrated that gender discrimination is less pervasive than ethnic discrimination. However, the gender bias appears to manifest selectively across occupations and work-hour requirements. Fictitious female candidates encounter lower interview invitation scores in roles traditionally dominated by men, such as sales representatives ($\lambda_{male}$ = 64.40, $CI_{95\%}$ = [63.89; 64.91]; $\lambda_{female}$ = 63.87, $CI_{95\%}$ = [62.20; 65.53]), site managers ($\lambda_{male}$ = 66.79, $CI_{95\%}$ = [65.49; 68.10]; $\lambda_{female}$ = 64.58, $CI_{95\%}$ = [64.02; 65.14]), and technical production managers ($\lambda_{male}$ = 59.61, $CI_{95\%}$ = [58.73; 60.48]; $\lambda_{female}$ = 58.95, $CI_{95\%}$ = [58.69; 59.21])—female shares in these occupations respectively approximate 10%, 29%, and 35% based on international data (International Labour Organization, 2024). In contrast, I observe bias against male candidates in the occupations of legal assistant ($\lambda_{male}$ = 65.81, $CI_{95\%}$ = [64.39; 67.24]; $\lambda_{female}$ = 67.17, $CI_{95\%}$ = [66.83; 67.50]) and seller of clothing accessories ($\lambda_{male}$ = 70.09, $CI_{95\%}$ = [68.91; 71.28]; $\lambda_{female}$ = 72.54, $CI_{95\%}$ = [71.15; 73.92]) with female occupational shares of 40% and 48%, and in part-time positions ($\lambda_{male}$ = 67.49, $CI_{95\%}$ = [65.87; 69.11]; $\lambda_{female}$ = 69.55, $CI_{95\%}$ = [67.24; 71.85]). Marginal effects at the mean by gender identity and job characteristics can be retrieved from Tables A17 to A24 and Figures A6 to A7 in the appendix.

These results align with observations in field experiments with human recruiters, where women are discriminated against in (higher-paying) occupations dominated by men and men are discriminated against in (lower-paying) occupations dominated by women (Galos & Coppock, 2023). They also indicate that the gender bias may be contextually linked to societal stereotypes or gendered expectations associated with particular occupations and work arrangements, likewise corresponding to a statistical discrimination mechanism.

### 3.3.3. Discrimination by sampling strategy

Another series of analyses concerns the correlation between ChatGPT's sampling strategy and its bias. The sampling strategy is included in ChatGPT's temperature parameter, which modulates the degree of randomness or 'creativity' in ChatGPT's output. Low temperatures (i.e. a deterministic sampling strategy) result in more coherent and consistent responses, while higher temperatures (i.e. a stochastic sampling strategy) produce more varied outputs. As the temperature increases, ChatGPT may diverge from its common pre-trained biases but could also introduce and reinforce uncommon biases. Using OLS regressions to estimate the interaction effects between candidate ethnic identity and model temperature, ChatGPT's sampling strategy appears to have no impact on its ethnic bias (see Table A25 in the appendix). In other words, making the sampling process more stochastic (i.e. introducing randomness) does not significantly change ChatGPT's bias. This finding hints that the uncovered bias arises from the pre-training data rather than being particular to the sampling during post-processing.

## 3.4. Discrimination at different cutoffs

Next, I evaluate whether selecting a proprietary cutoff score as an end decision-maker could impact the sign or strength of the ethnic bias identified in Section 3.2.1. In other words, does the decision of an end user, relying on ChatGPT's ratings, to invite a set of candidates who attain a particular minimum score influence the extent to which they would discriminate in the selection process?

Selecting a proprietary cutoff score does not change the sign of the resulting ethnic bias but does change its magnitude. Panel A of Figure 4 illustrates that at every cutoff score in [50,80], Dutch applicants receive higher invitation probabilities on average than candidates with other ethnic identities. For example, at a threshold value of 60, Dutch applicants would be invited 74.92% of the time, on average, compared to 69.20% for the group of other candidates—i.e. a difference of about 5.7 percentage points. This discrepancy equates to a discrimination ratio of 0.92. Ratios below 1 indicate discrimination against candidates with a different ethnic identity, with lower ratios implying a higher discrimination level (see

Section 2.3.4 for calculation details). More specifically, applicants of other ethnic identities than Dutch would be invited 8% less frequently.

The absolute gap tightens in the [70,80) range. At a cutoff score of 75, for example, the difference amounts to 2.8 percentage points; Dutch applicants would be invited 18.79% of the time on average, while the average invitation probability of other candidates at this threshold equals 16.00%. However, in relative terms, this difference equates to a discrimination ratio of 0.85, indicating a 15% lower relative invitation probability and, thus, more discrimination against other ethnicity applicants compared to a cutoff score of 60. For cutoff scores in [1,50) and [80,100], the outcome is nearly the same for all applicants, namely that virtually all or no candidates would be invited, respectively. Decision-makers using ChatGPT as an assistant for personnel selection would logically only choose a cutoff in [50,80)—score differentiation only occurs in this range—and, consequently, discriminate.

Panel B of Figure 4 zooms in on the results by ethnic identity in the [50,80) range. The results are similar to the findings based on the OLS model estimates: candidates with White American names are penalised the least compared to Dutch applicants. In contrast, Asian- and Eastern European-named applicants face the largest penalties. The most considerable difference in interview invitation probability in the [50,80) range occurs around a threshold value of 75. At this score, discrimination ratios amount to 0.81 for Asians, 0.83 for Eastern Europeans and Hispanics, 0.84 for Black Americans, 0.85 for Arabs, 0.86 for Turks and Central Africans, and 0.93 for White Americans. To put these in perspective, average discrimination ratios of hiring discrimination by humans in worldwide audit research equal approximately 0.67 for Asians, 0.72 for Eastern Europeans, 0.92 for Hispanics, 0.68 for Blacks, 0.59 for Arabs, and 0.75 for Western Asians including Turks (Lippens, Vermeiren, & Baert, 2023). Average discrimination ratios in Belgium based on the most recent experimental evidence from correspondence audits approximate 0.73 for Eastern Europeans, 1.00 for Blacks, 0.79 for Arabs, and 0.85 for Turks (Lippens, Dalle, D'hondt, Verhaeghe, & Baert, 2023). Even though ChatGPT discriminates based on ethnic identity, the above observations suggest that it performs better than the average human recruiter (except for the Hispanic subgroup measured in audits worldwide and the Black subgroup measured in Belgian (Dutch) research). However, note that substantial differences in name sets and control groups between the estimates presented in the current study, with its specific experimental setup, and the average estimates from audit research with humans make a formal comparison difficult.

# 4. Conclusion

Through a simulated CV screening task based on the correspondence audit approach, I provide evidence that ChatGPT displays systematic bias in its output, showing noticeable preferences for specific ethnic and gender groups when evaluating job candidates. The chatbot was significantly less inclined to advance equally qualified Arabs, Asians, Black Americans, Central Africans, Eastern Europeans, Hispanics, Turks, and White Americans to the interview stage of the selection process than candidates from the Dutch reference group. The minor penalty for White American-named and more substantial penalty for Black American-named candidates suggest that a prompt language-specific bias is at play alongside an ethnic-specific bias. Levels of ethnic identity bias by ChatGPT appeared lower compared with meta-analytic ethnic hiring discrimination estimates from worldwide correspondence audit research involving human recruiters—although it is essential to highlight the differences in name sets and control groups between this experiment's setup and correspondence audits with humans in existing field research. Moreover, female candidates were not rated lower on average than their male counterparts. At the intersection of ethnic and gender identity, however, ChatGPT discriminated more against Turkish females than Turkish males, consistent with the double minority status of the former group.

The heterogeneity analyses further elucidated a job context-dependent bias. ChatGPT's ethnic bias is particularly pronounced in roles offering favourable labour conditions—full-time contracts and day shift systems—and roles with greater language proficiency requirements. Gender bias, conversely, surfaces in gender-atypical jobs; female candidates are discriminated against in male-dominated occupations, whereas male candidates are discriminated against in female-dominated occupations and part-time vacancies. The LLM's discrimination mechanism appears to align with statistical discrimination, where ChatGPT relies on group characteristics and productivity signals from its pre-training data, reproducing societal stereotypes. Altering the chatbot's sampling strategy did not significantly influence its bias. Finally, I observed substantial heterogeneity in interview scores between names of the same ethnic and gender identity.

The potential risks associated with integrating LLMs in selection decisions hinge highly on who uses the technology, how, and when rather than on the inherent qualities of LLMs. While the findings from this study indicate that ChatGPT exhibits a level of bias generally falling behind that of human recruiters, I can envisage a scenario where its use leads to an increase in selection bias. For example, if ChatGPT is deployed for automated pre-screening, bias in its output may be amplified by human prejudice in subsequent manual screening stages. This sequence of events can result in an overall more discriminatory selection process. Such a scenario highlights that reliance on LLMs for selection can inadvertently perpetuate or aggravate discriminatory treatment even if the language model itself is less discriminatory than a human screener, resulting in cumulative discrimination.

I see several avenues for future research. First, scholars could explore whether the uncovered bias transposes to different grounds for discrimination, languages, contexts, selection tasks, or language models. For example, discrimination based on age is a persistent problem in personnel selection, which also merits attention in research on AI bias (Lippens, Vermeiren, & Baert, 2023; Stypinska, 2023). AI-based decision-making may also result in discrimination in contexts such as housing and healthcare through algorithm-based awarding of house rentals or treatment plan recommendations in patients, to name just two examples (Basu, 2023; Rosen et al., 2021). Second, to enhance model explainability, researchers could continue investing in making the decision-making processes of large language models more transparent to understand which features contribute to the bias in their output (Arrieta et al., 2020). Third, to reduce bias, scholars could further evaluate fairness-enhancing techniques applied to the training data, the training process, or the post-processing (Friedler et al., 2019). One example of a bias-reducing technique applied to the prompt is anonymising personal information by removing explicit minority status markers before parsing. This technique has been proven effective in countering human prejudice (Åslund & Skans, 2012; Blommaert & Coenders, 2023; Derous & Ryan, 2019; Lacroux & Martin-Lacroux, 2019). Nonetheless, LLMs could still pick up on implicit markers in applicant profiles (e.g. organisation affiliations signalling ethnic group membership or years of professional experience signalling age), continuing the output of biased responses (Arnold et al., 2021; Kleinberg et al., 2018).

This study underlines the significance of understanding and addressing systemic bias in large language models, especially when deployed in real-world applications such as hiring and selection. Three parties can help in this endeavour. First, model developers preferably advance efforts to mitigate biases arising from the pre-training data and process. Second, policymakers may create legal frameworks ensuring equitable use of large language models in decision-making directly impacting humans. The AI Act recently agreed upon by the European Parliament and European Council forms a step in the right direction (European Commission, 2021). The act bounds LLMs by imposing transparency obligations and restrictions on their use in automatic categorisation and selection, which carry actual potential risks, as illustrated in this study. Third, practitioners considering using ChatGPT and the like should proceed cautiously. At a minimum, their usage requires a thorough assessment of the trade-off between increased efficiency and adverse impact. Taken together, the applicability of LLMs in their current form for activities that involve decision-making affecting humans is debatable.

# Declarations

## Data and code availability

The data and code used in this study, as well as supplementary tables and figures, are available at https://osf.io/vezt7/ (licensed under CC-BY-4.0).

## Declaration of competing interest

I declare no relevant financial or non-financial competing interests.

## Declaration of generative AI

While preparing this work, I used DeepL, Grammarly, and ChatGPT 4 to improve the readability and language of some of the writing. After using these tools, I reviewed, edited, and paraphrased the content as needed. I take full final responsibility for the content of the publication.

## Funding

This study was conducted on the margins of the EdisTools project. EdisTools is funded by FWO (Research Foundation – Flanders) under grant number S004119N. I also acknowledge funding from FWO under grant number 12AM824N.

## Author contributions

L.L. was the sole contributor to this manuscript.

# Tables

Table 1. OLS regression of ChatGPT interview invitation scores on ethnic and gender identity

| | (1) | (2) | (3) | (4) | (5) | (6) | (7) |
|---|---|---|---|---|---|---|---|
| Intercept | 67.2157*** | 67.1807*** | 69.0142*** | 69.0244*** | 69.0422*** | 70.7160*** | 70.4346*** |
| | (0.2571) | (0.2585) | (2.5631) | (2.4994) | (1.7973) | (2.1083) | (2.0392) |
| **Panel A. Identity (ref.: Dutch)** | | | | | | | |
| White American | −0.9563*** | −0.9563*** | −0.9562*** | −0.9562*** | −0.9567*** | −0.9566*** | −0.9569*** |
| | (0.2258) | (0.2244) | (0.1861) | (0.1891) | (0.1900) | (0.1899) | (0.1880) |
| Arab | −1.4117*** | −1.4117*** | −1.4124*** | −1.4125*** | −1.4129*** | −1.4125*** | −1.4123*** |
| | (0.2141) | (0.2192) | (0.1995) | (0.2014) | (0.1991) | (0.1966) | (0.1949) |
| Central African | −1.5468*** | −1.5468*** | −1.5474*** | −1.5475*** | −1.5467*** | −1.5466*** | −1.5469*** |
| | (0.2278) | (0.2378) | (0.2439) | (0.2411) | (0.2345) | (0.2359) | (0.2359) |
| Hispanic | −1.6257*** | −1.6257*** | −1.6255*** | −1.6257*** | −1.6247*** | −1.6253*** | −1.6255*** |
| | (0.2288) | (0.2378) | (0.2275) | (0.2345) | (0.2334) | (0.2301) | (0.2279) |
| Turkish | −1.7478*** | −1.7478*** | −1.7470*** | −1.7469*** | −1.7470*** | −1.7472*** | −1.7473*** |
| | (0.2227) | (0.2258) | (0.2152) | (0.2187) | (0.2175) | (0.2074) | (0.2113) |
| Black American | −1.8436*** | −1.8436*** | −1.8424*** | −1.8423*** | −1.8420*** | −1.8426*** | −1.8429*** |
| | (0.2288) | (0.2302) | (0.2431) | (0.2475) | (0.2502) | (0.2477) | (0.2497) |
| Asian | −2.1583*** | −2.1583*** | −2.1586*** | −2.1585*** | −2.1579*** | −2.1576*** | −2.1579*** |
| | (0.2340) | (0.2359) | (0.2773) | (0.2843) | (0.2795) | (0.2776) | (0.2808) |
| Eastern European | −2.4170*** | −2.4170*** | −2.4168*** | −2.4168*** | −2.4169*** | −2.4170*** | −2.4170*** |
| | (0.2275) | (0.2324) | (0.2389) | (0.2412) | (0.2467) | (0.2418) | (0.2479) |
| **Panel B. Gender (ref.: Male)** | | | | | | | |
| Female | | 0.0705 | 0.0707 | 0.0707 | 0.0711 | 0.0709 | 0.0707 |
| | | (0.1530) | (0.1504) | (0.1526) | (0.1481) | (0.1511) | (0.1515) |
| **Panel C. Covariates** | | | | | | | |
| ChatGPT sampling strategy | Yes | Yes | Yes | Yes | Yes | Yes | Yes |
| Candidate occupation | No | No | Yes | Yes | Yes | Yes | Yes |
| Job location | No | No | No | Yes | Yes | Yes | Yes |
| Job experience | No | No | No | No | Yes | Yes | Yes |
| Job language requirements | No | No | No | No | No | Yes | Yes |
| Job type, shift system, & work hours | No | No | No | No | No | No | Yes |
| **Panel D. Model parameters** | | | | | | | |
| N | 34,560 | 34,560 | 34,560 | 34,560 | 34,560 | 34,560 | 34,560 |
| $R^2$ | 0.005 | 0.005 | 0.088 | 0.091 | 0.144 | 0.163 | 0.168 |
| $R^2$ Adj. | 0.005 | 0.005 | 0.087 | 0.090 | 0.143 | 0.161 | 0.166 |
| AIC | 276,647 | 276,649 | 273,697 | 273,605 | 271,530 | 270,774 | 270,595 |
| BIC | 276,782 | 276,793 | 274,027 | 274,019 | 271,961 | 271,298 | 271,220 |

*Notes.* Abbreviations and acronyms used: ref. (reference group), N (sample size), Adj. (Adjusted), AIC (Akaike information criterion), and BIC (Bayesian information criterion). Model statistics are OLS coefficient estimates with standard errors between parentheses. Standard errors are corrected using cluster-robust wild bootstrapping with 2,000 replications. Clusters are defined at the vacancy level, given the correlation between the assignment of the candidates and the vacancies. * $p < .05$; ** $p < .01$; *** $p < .001$.

Table 2. OLS estimation of the moderation effect of gender on ChatGPT's ethnic bias

| | (1) | (2) |
|---|---|---|
| Intercept | 66.9115*** (0.2993) | 72.5865*** (1.0395) |
| **Panel A. Identity (ref.: Dutch)** | | |
| White American | −1.1338*** (0.3198) | −1.1346*** (0.3159) |
| Arab | −1.6368*** (0.3213) | −1.6376*** (0.3200) |
| Central African | −1.3688*** (0.3305) | −1.3690*** (0.3237) |
| Hispanic | −1.3106*** (0.3220) | −1.3104*** (0.3329) |
| Turkish | −0.8595** (0.3243) | −0.8587* (0.3356) |
| Black American | −1.3687*** (0.3180) | −1.3679*** (0.3312) |
| Asian | −1.7234*** (0.3305) | −1.7234*** (0.3386) |
| Eastern European | −1.8941*** (0.3286) | −1.8954*** (0.3311) |
| **Panel B. Gender (ref.: Male)** | | |
| Female | 0.6063 (0.3221) | 0.6061 (0.3207) |
| **Panel C. Identity x Gender (ref.: Dutch x Male)** | | |
| White American x Female | 0.3551 (0.4598) | 0.3555 (0.4617) |
| Arab x Female | 0.4504 (0.4490) | 0.4506 (0.4438) |
| Central African x Female | −0.3560 (0.4572) | −0.3559 (0.4553) |
| Hispanic x Female | −0.6303 (0.4478) | −0.6303 (0.4603) |
| Turkish x Female | −1.7765*** (0.4810) | −1.7771*** (0.4794) |
| Black American x Female | −0.9499* (0.4753) | −0.9503* (0.4830) |
| Asian x Female | −0.8696 (0.4586) | −0.8690 (0.4566) |
| Eastern European x Female | −1.0459* (0.4767) | −1.0432* (0.4597) |
| **Panel D. Covariates** | | |
| ChatGPT sampling strategy | Yes | Yes |
| Candidate occupation | No | Yes |
| Job location | No | Yes |
| Job experience | No | Yes |
| Job language requirements | No | Yes |
| Job type, shift system, & work hours | No | Yes |
| **Panel E. Model parameters** | | |
| N | 34,560 | 34,560 |
| $R^2$ | 0.006 | 0.169 |
| $R^2$ Adj. | 0.005 | 0.167 |
| AIC | 276,642 | 270,584 |
| BIC | 276,853 | 271,277 |

*Notes.* Abbreviations and acronyms used: ref. (reference group), N (sample size), Adj. (Adjusted), AIC (Akaike information criterion), and BIC (Bayesian information criterion). Model statistics are OLS coefficient estimates with standard errors between parentheses. Standard errors are corrected using cluster-robust wild bootstrapping with 2,000 replications. Clusters are defined at the vacancy level, given the correlation between the assignment of the candidates and the vacancies. * $p < .05$; ** $p < .01$; *** $p < .001$.

# Figures

Figure 1. Histogram of ChatGPT invitation scores by ethnic identity

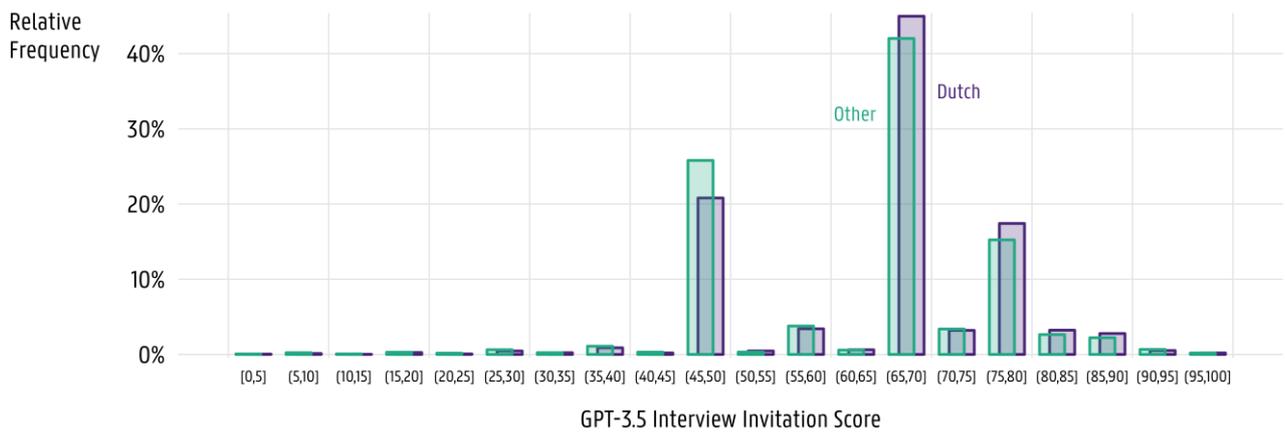

*Notes.* Green (light) bars represent binned relative frequencies of fictitious Arab-, Asian-, Black American-, Central African-, Eastern European-, Hispanic-, Turkish-, and White American-named candidates used in the simulated CV screening task. Purple (dark) bars represent those of Dutch-named applicants.

Figure 2. Average (predicted) ChatGPT interview invitation scores by ethnic identity

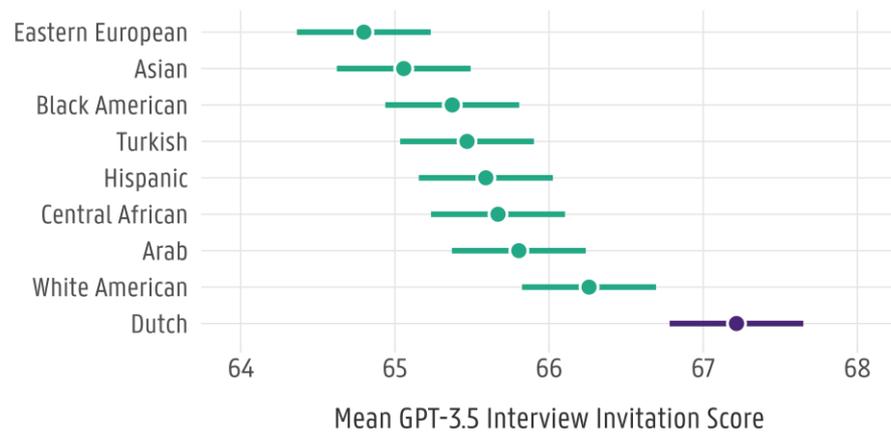

*Notes.* Points are predicted values or estimated marginal effects at the mean. Lines illustrate the 95% confidence intervals of these values. Estimates are based on Model 1 in Table 1.

Figure 3. Average ChatGPT interview invitation scores by name, ethnic and gender identity

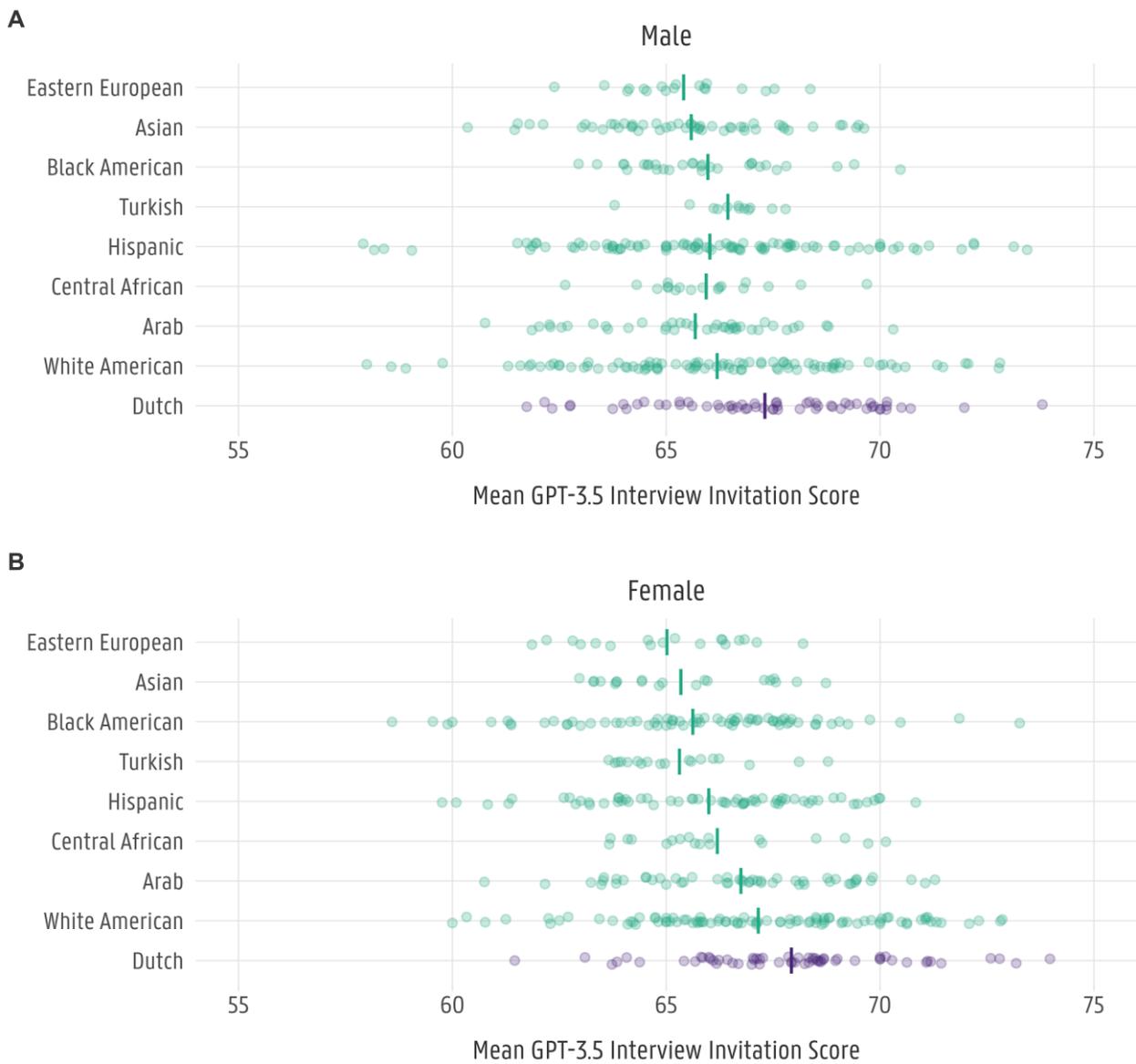

*Notes.* Points represent mean interview invitation scores for each name across ethnic identities and genders. These points are slightly vertically jittered to more easily discern distribution densities. Vertical lines depict score means by ethnic and gender identity. Panel A includes male names, whereas Panel B includes female names.

Figure 4. Average (predicted) ChatGPT interview invitation probabilities by ethnic identity

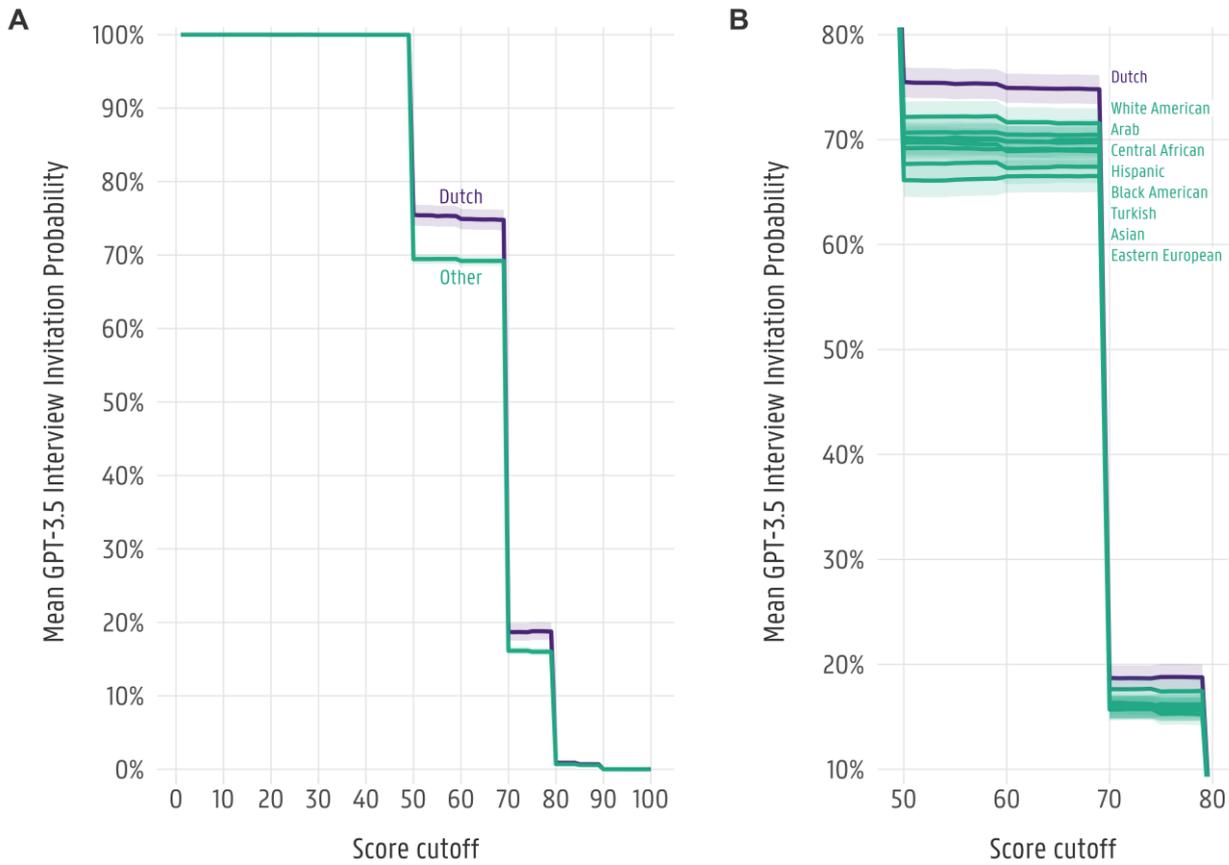

*Notes*. Values along the curves represent predicted interview probabilities or estimated marginal effects at the mean of candidates with a given ethnic identity at each possible cutoff score in [1,100] ∩ ℕ based on ChatGPT's output. Estimates are derived from the logit model specification in Equation 6 (see Section 2.3.4). Lightly shaded ribbons illustrate the 95% confidence intervals of these estimates. Panel A depicts the differences in predicted probabilities between Dutch candidates and candidates with another ethnic identity for all cutoff scores in [1,100] ∩ ℕ. Panel B zooms in on cutoff scores in [50,80] and shows the interview probabilities for each ethnic identity separately. Detailed results of the separate logit models are available from the OSF files associated with this paper (https://osf.io/vezt7/).

# Appendix

Supplementary tables and figures are available at https://osf.io/vezt7/ (licensed under CC-BY-4.0).